\documentstyle[12pt]{article}
\newcommand{\be}{\begin{equation}}
\newcommand{\ee}{\end{equation}}
\newcommand{\bea}{\begin{eqnarray}}
\newcommand{\eea}{\end{eqnarray}}
\newcommand{\wt}{\widetilde}
\newcommand{\wh}{\widehat}
\def\vu{\varepsilon}

\def\a{\alpha}
\def\b{\beta}
\def\d{\delta}

\def\f{\phi}

\def\h{\eta}
\def\k{\kappa}

\def\q{\theta}

\def\s{\sigma}

\def\F{\Phi}
\def\G{\Gamma}

\newcommand{\nonu}{\nonumber \\[2mm]}
\newcommand{\del}{\partial}
\newcommand{\delb}{\bar{\partial}}
\begin{document}
\renewcommand{\theequation}{\thesection.\arabic{equation}}
\newcommand{\eqn}[1]{eq.(\ref{#1})}
\renewcommand{\section}[1]{\addtocounter{section}{1}
\vspace{5mm} \par \noindent
  {\bf \thesection . #1}\setcounter{subsection}{0}
  \par
   \vspace{2mm} } 
\newcommand{\sectionsub}[1]{\addtocounter{section}{1}
\vspace{5mm} \par \noindent
  {\bf \thesection . #1}\setcounter{subsection}{0}\par}
\renewcommand{\subsection}[1]{\addtocounter{subsection}{1}
\vspace{2.5mm}\par\noindent {\em \thesubsection . #1}\par
 \vspace{0.5mm} }
\renewcommand{\thebibliography}[1]{ {\vspace{5mm}\par \noindent{\bf
References}\par \vspace{2mm}}
\list
 {\arabic{enumi}.}{\settowidth\labelwidth{[#1]}\leftmargin\labelwidth
 \advance\leftmargin\labelsep\addtolength{\topsep}{-4em}
 \usecounter{enumi}}
 \def\newblock{\hskip .11em plus .33em minus .07em}
 \sloppy\clubpenalty4000\widowpenalty4000
 \sfcode`\.=1000\relax \setlength{\itemsep}{-0.4em} }
{\hfill{VUB/TENA/96/06, hep-th/9610102}}
\vspace{1cm}
\begin{center}
{\bf OFF-SHELL FORMULATION OF N=2 NON-LINEAR $\sigma$-MODELS }

\vspace{1.4cm}

ALEXANDER SEVRIN and JAN TROOST\footnote{Aspirant NFWO} \\
{\em Theoretische Natuurkunde, Vrije Universiteit Brussel} \\
{\em Pleinlaan 2, B-1050 Brussel, Belgium} \\
\end{center}
\centerline{ABSTRACT}

\vspace{- 4 mm}  

\begin{quote}\small
We study $d=2$, $N=(2,2)$ non-linear $\sigma$-models in $(2,2)$
superspace. By analyzing the most general constraints on a superfield, we show
that through an appropriate choice of coordinates, there are
no other superfields than chiral,
twisted chiral and semi-chiral ones. We study
the resulting $\sigma$-models and we speculate on the possibility that
{\it all} $(2,2)$ non-linear $\sigma$-models
can be described using these fields. We apply the results to two examples:
the $SU(2)\times U(1)$ and the $SU(2)\times SU(2)$ WZW model. Pending upon
the choice of complex structures,
the former
can be described in terms of either one semi-chiral multiplet or a chiral
and a twisted chiral multiplet.
The latter is formulated in
terms of one semi-chiral and one twisted chiral multiplet. For both cases
we obtain the potential explicitely.
\end{quote}
\baselineskip17pt
\addtocounter{section}{1}
\par \noindent
  {\bf \thesection . Introduction}
  \par
   \vspace{2mm} 
\noindent
Supersymmetric non-linear $\sigma$-models, in particular those with two or
more supersymmetries, are building blocks for stringtheories. In
\cite{luis} torsion free supersymmetric non-linear $\sigma$-models were
studied and this resulted in a full classification of models within this
class:
$N=1$ is always possible, $N=2$ is in one-to-one correspondence
with the target manifold being K\"ahler, $N=3$ implies $N=4$, $N=4$ requires
the target
manifold to be hyper-K\"ahler and $N>4$ is not possible. Later \cite{tor},
$\sigma$-models which do have torsion were studied but, except for the fact
that $N=1$
is always possible, no general
classification has been given. In \cite{usold} a subclass of
torsionful $\sigma$-models were studied: the Wess-Zumino-Witten models,
where the targetmanifold is a semi-simple Lie groupmanifold. There the
classification of extended supersymmetries has been done: again $N\leq 4$ and
$N=2$ is possible
on all even-dimensional groups, $N=3\Rightarrow N=4$,
$N=4$ is only possible on
those groups which could be decomposed as ``products'' of Wolf
spaces.

A second problem arises when one considers supersymmetry
transformations of both chiralities, where one
finds that the
supersymmetry algebra closes only {\it on-shell}. Finding auxiliary fields
such that the supersymmetry algebra closes off-shell, {\it i.e.} gets
realized in a model independent way, is equivalent to finding a
superspace formulation for these models. Both the $N=(1,1)$ and
$N=(2,1)$, \cite{DS}, cases are easily
solved in their full generality while for torsionful models with
$N\geq (2,2)$ this remains largely
unsolved. Having an extended superspace description of these models yields
various simplifications and applications. The geometry becomes transparant
and duality transformations can be performed which keep the extended
supersymmetry manifest \cite{RSS,kiritsis}.

In this paper we study the
$N=(2,2)$ case. The action in $(2,2)$ superspace is just the integral of a
potential, so in this way one might get a handle on the geometry of $N=(2,2)$
supersymmetric non-linear $\sigma$-models. As the lagrange density is a
potential, the dynamics is now also determined by the choice of
superfields. Till now several types of superfields were discovered. The
simplest case corresponds to models formulated solely in terms of chiral
fields. These $\sigma$-models are K\"ahler. A generalization thereof is
obtained by using both chiral and twisted chiral superfields \cite{GHR}. The
resulting model is then a mild generalization of the K\"ahler case. These
models do have torsion. A last type of superfields are the semi-chiral fields
\cite{BLR}.
Though discovered several years ago, they have barely been studied. As
will become clear in this paper, models which include semi-chiral
superfields represent a radical departure of the two previously mentioned
types of $\sigma$-models.

In this paper we show by analyzing the most general constraints one can
impose on superfields, that the previously mentioned three types of fields
is all there is. Subsequently we study the most general model and provide
several arguments which support the hope that {\it all} $N=(2,2)$
non-linear $\sigma$-models can be described using these fields.
Finally
we give two explicit examples: $SU(2)\times U(1)$ and $SU(2)\times SU(2)$,
the former being described either by a single semi-chiral multiplet or
by a chiral and a twisted chiral multiplet and the
latter by one semi-chiral and one chiral multiplet.
\setcounter{equation}{0}
\section{Supersymmetric non-linear $\sigma$-models}
Omitting the dilaton term, a general supersymmetric
non-linear $\sigma$-model in $N=(1,1)$ superspace is given by\footnote{
We take $D\equiv\frac{\del}{\del \q}+\q\del$ and
$\bar D \equiv \frac{\del}{\del \bar \q}+\bar\q\delb$,
with $\partial\equiv\frac{\partial}{\partial z}$ and
$\bar\partial\equiv\frac{\partial}{\partial{\bar z}}$.}
\be
{\cal S}=\int d^2 x d^2 \q\left(g_{ab}+b_{ab}\right)D\f^a\bar D
\f^b,\label{staction}
\ee
The metric on the target manifold is $g_{ab}$ while $b_{ab}=-b_{ba}$ is a
potential for the
torsion $T$,
\be
T^a{}_{bc}\equiv-\frac 3 2 g^{ad}b_{[bc,d]}.\label{Sdef}
\ee
The equations of motion follow from eq. (\ref{staction}):
\bea
D\bar D \f^a + \G^a_{-cb}  D\f^b\bar D
\f^c=0,
\eea
where we used one of the two natural connections $\Gamma_\pm$:
\be
\G^a_{\pm bc}\equiv  \left\{ {}^{\, a}_{bc} \right\}
\pm T^a{}_{bc},
\ee
the first term being the standard Levi-Cevita connection.
The action is invariant under the supersymmetry transformations
\be
\d\f^a = \vu Q \f^a +\bar\vu\bar Q\f^a,\label{sus1}
\ee
where
\be
Q\equiv\frac{\del}{\del \q}-\q\del,\qquad
\bar Q \equiv \frac{\del}{\del \bar \q}-\bar\q\delb .
\ee
The commutator of two supersymmetries closes on a translation
\be
\left[ \d_1 , \d_2 \right]\f^a = 2 \vu_2\vu_1\del\f^a + 2
\bar\vu_2\bar\vu_1\delb\f^a.\label{suscom}
\ee
A second supersymmetry is necessarily of the form
\be
\d \f^a=\h J^a{}_bD\f^b+\bar\h\bar J^a{}_b\bar D\f^b\label{sus2}.
\ee
The action, eq. (\ref{staction}), is invariant under the
transformations eq. (\ref{sus2}) provided\footnote{By $\nabla^\pm$ we denote
covariant differentiation using the  $\Gamma_\pm$ connection.}
\be
\nabla_c^+J^a{}_b=\nabla_c^-\bar J^a{}_b=0,\qquad
g_{bc}J^c{}_a=-g_{ac}J^c{}_b,\qquad g_{bc}\bar J^c{}_a=-g_{ac}\bar
J^c{}_b, \label{herm}
\ee
hold.
One obtains the standard {\it on-shell} supersymmetry algebra, {\it i.e.} the
first
and the second supersymmetry commute, and the second supersymmetry
satisfies the same algebra as the first, if $J$ and $\bar J$ obey
\begin{eqnarray}
J^2=\bar J^2=-{\bf 1},\qquad N^a{}_{bc}[J,J]=N^a{}_{bc}[\bar J,\bar J]=0,
\end{eqnarray}
with the Nijenhuis tensor, $N[A,B]$, given by
\be
N^a{}_{bc}\left[A ,B\right]\equiv A^d{}_{[b}B^a{}_{c],d}+A^a{}_dB^d{}_{[b,c]}+
B^d{}_{[b}A^a{}_{c],d}+B^a{}_dA^d{}_{[b,c]}.
\label{Nij2}
\ee
In other words the model is $(2,2)$ supersymmetric iff. the manifold allows
for two complex structures for both of which the metric is hermitean. Each
of these complex structures is covariantly constant, however w.r.t.
different connections. Note that though each of the complex structures is
{\it individually} integrable, they are not necessarily {\it simultanously}
integrable.

A few more interesting formulae can be obtained.
Using the constancy of the complex structures, we can rewrite the
Nijenhuis condition as
\begin{eqnarray}
3T_{ef[a}J^e{}_bJ^f{}_{c]}=T_{abc},\label{r1}
\end{eqnarray}
and similarly for $\bar J$. Another consequence of the constancy of the
complex structures is:
\begin{eqnarray}
J_{[ab,c]}=2T^d{}_{[ab}J_{c]d}, \qquad
\bar{J}_{[ab,c]}=-2T^d{}_{[ab}\bar{J}_{c]d},
\label{kahler}
\end{eqnarray}
which gives the exterior derivative of the two fundamental two-forms
associated with $J$ and $\bar J$.
Combining eqs. (\ref{kahler}) and (\ref{r1}) results in an expression for the
torsion in terms of the complex structures:
\begin{eqnarray}
T_{abc}= \frac 3 2 J^d{}_a J^e{}_b J^f{}_c J_{[de,f]}=
-\frac 3 2 \bar{J}^d{}_a \bar{J}^e{}_b \bar{J}^f{}_c \bar{J}_{[de,f]}.
\label{torJ}
\end{eqnarray}
The supersymmetry algebra is standard but for the commutator of the
left-handed second supersymmetry with the right handed second one:
\begin{eqnarray}
[\delta(\eta ),\delta (\bar\eta)]\phi^a = \eta\bar\eta[J,\bar J]^a{}_b(
D\bar D \f^b + \G^b_{-cd}  D\f^d\bar D \phi^c).
\end{eqnarray}
So as long as the left and right complex structure commute, the supersymmetry
algebra closes off-shell and  we expect that
we can formulate the model in (2,2) superspace without introducing any new
fields. However, if they do not commute, the supersymmetry algebra closes only
on-shell,
hence the algebra is model dependent and a manifest (2,2)
supersymmetric formulation will require the introduction of additional
auxiliary fields.

Consider now the case where $[J,\bar J]=0$. In the appendix we show that
$N[J,J]=N[\bar J,\bar J]=0$ imply that $N[J,\bar J]= N[J,\Pi]=N[\bar
J,\Pi]=N[\Pi,\Pi]=0$, where $\Pi\equiv J \bar J$ is a product structure,
\cite{GHR}:
$\Pi^2={\bf 1}$. As was to be expected, the fact that the complex
structures commute, guarantees that they are simultanously integrable.
The product structure allows us to introduce projection operators:
\begin{eqnarray}
P_\pm\equiv \frac 1 2 (1\pm \Pi).
\end{eqnarray}
{}From
\begin{eqnarray}
\ker\, [J,\bar J]=\ker(J+\bar J) \oplus  \ker(J-\bar J),\label{ker}
\end{eqnarray}
we get that $P_+$ projects on $\ker(J+\bar J)$, and $P_-$ on $\ker(J-\bar J)$.
If we have that
$J=\pm \bar J$, then it follows from eq. (\ref{kahler}) that
the fundamental two form
is closed and eq. (\ref{torJ}) implies
that the torsion vanishes: the manifold is K\"ahler \cite{luis}.
Having $J\neq \pm \bar J$, gives a manifold with a product structure. The
subspaces obtained by the projection operators $P_\pm$ are K\"ahler and
the manifold has torsion given by eq. (\ref{torJ}). We come back to this
in the next section. Conversely, one can also show that any model for
which $[J,\bar J]= 0$ can be described by chiral and twisted chiral fields
\cite{GHR}.
One example of a manifold which can be described by such a ``twisted K\"ahler''
geometry
is
$SU(2) \times U(1)$ \cite{RSS}.
\setcounter{equation}{0}
\section{$N=(2,2)$ superfields}
\subsection{Chiral and twisted chiral superfields}
\noindent
We now turn to a manifest $(2,2)$ supersymmetric formulation of
non-linear $\sigma$-models. To achieve this we introduce $(2,2)$
superspace which conists of two fermionic directions for
each chirality, hence the name.
Its left-handed (right-handed) sector
is parametrized by the coordinates $(z,\theta^+,\theta^-)$
($(\bar z,\bar \theta^+,\bar \theta^-)$).
Covariant derivatives are given by
\begin{eqnarray}
D_\pm \equiv\frac{\partial}{\partial\theta^\pm }+\theta^\mp \partial,
\qquad
\bar D_\pm \equiv\frac{\partial}{\partial\bar\theta^\pm }
+\bar\theta^\mp \bar\partial .   \label{defD}
\end{eqnarray}
Introducing
\be
Q_\pm=\frac{\partial}{\partial\theta^\pm}-\q^\mp\del; \qquad\quad
\bar Q_\pm=\frac{\partial}{\partial\bar \theta^\pm}-\bar
\q^\mp\delb\, ,
\ee
we get the $N=(2,2)$ supersymmetry transformation on a general superfield
$\Phi$:
\be
\d \F = \vu^+Q_+\F+ \vu^-Q_-\F+\bar  \vu^+\bar Q_+\F+ \bar \vu^-\bar
Q_-\F.
\ee
Passing from $(1,1)$ to $(2,2)$ superspace is facilitated through the use
of the original fermionic coordinates $\theta$ and $\bar\theta$:
\be
\q = \frac{1}{\sqrt{2}}(\q^++\q^-),\qquad \bar \q = \frac{1}{\sqrt{2}}
(\bar\q^++\bar\q^-),
\ee
and the extra fermionic coordinates
\be
\wh\q = \frac{1}{\sqrt{2}}(\q^+-\q^-),\qquad \wh{\bar \q} =
\frac{1}{\sqrt{2}} (\bar\q^+-\bar\q^-).
\ee
In this way we get {\it e.g.}
\be
D=\frac{1}{\sqrt{2}}(D_++D_-)=\frac{\del}{\del \q}+\q\del,\qquad
\wh D=\frac{1}{\sqrt{2}}(D_+-D_-)=\frac{\del}{\del \wh \q}-\wh
\q\del ,\qquad
\ee
and similarly for $\bar D$ and $\wh{\bar D}$. The supersymmetry generators
are
\be
Q\equiv \frac{1}{\sqrt{2}}( Q_+ + Q_-) = \frac{\del}{\del \q}-\q\del
\qquad   \wh Q=\frac{1}{\sqrt{2}}(Q_+-Q_-)=\frac{\del}{\del \wh
\q}+\wh \q\del,
\ee
and similar expressions for  $\bar Q$ and $\wh{\bar Q}$.
In particular one gets then that when passing to $N=(1,1)$ superspace $\hat
Q=\wh D$ and
$\hat{\bar Q}=\wh{\bar D}$.

On dimensional grounds we know that the Lagrange density has to be a
function of scalar fields. The dynamics is then largely determined by
the superfields we use or said in a different way, our choice of
representations.

Starting from a set of general $(2,2)$ superfields $\Phi^a$, $a\in
\{1,\cdots, d\}$,  we can impose constraints of the form\footnote{For
given l.h.s., we get that Lorentz invariance and counting dimensions, does
give eq. (\ref{con1}) as the most general constraint.}
\begin{eqnarray}
\wh{ D }\Phi^a=i J^a{}_b D\Phi^b , \label{con1}
\end{eqnarray}
with $J^a{}_b$, some $(1,1)$ tensor. This eliminates half of the degrees
of freedom of the superfields. However some integrability conditions have
to be met.
Computing $\wh{D}^2\Phi^a$ using the r.h.s. of previous equation, we get:
\begin{eqnarray}
\wh{D}^2\Phi^a=(J^2)^a{}_b\partial\Phi^b+\frac 1 2
N^a{}_{bc}[J,J]D\Phi^cD\Phi^b,
\end{eqnarray}
which  is consistent with
$\wh D ^2=-\partial$ iff. $J^2=-1$ and the Nijenhuis tensor vanishes, hence
$J$ needs to be a complex structure.

Additional constraints of opposite chirality can be imposed and
have the form
\begin{eqnarray}
\wh{ \bar D }\Phi^a=i \bar{J}(\Phi)^a{}_b \bar D\Phi^b .  \label{con2}
\end{eqnarray}
Writing $\Phi^a$ as an expansion in $\hat \theta$ and $\hat {\bar\theta}$,
one verifies that eqs. (\ref{con1}) and (\ref{con2}) eliminate three of the
four components of  $\Phi^a$, thus effectively reducing $\Phi^a$ to a $(1,1)$
superfield.
Consistency of the constraint (\ref{con2})
with  $\wh {\bar D}^2=-\bar\partial$ requires $\bar J$ to
be a complex structure but $\{ \wh{ D}, \wh{\bar D}\}=0$ yields additional
conditions:
\begin{eqnarray}
\{ \wh{ D}, \wh{\bar D}\}\Phi^a= [\bar{J},J]^a{}_b\bar{D} D
\Phi^a+M^a{}_{bc}[J,\bar J] \bar D\Phi^cD\Phi^b.\label{ddbar}
\end{eqnarray}
The $(1,2)$ tensor $M^a{}_{bc}[J,\bar J]$ is discussed in the appendix,
where we show that as long as $[J,\bar J]=0$,
the vanishing of $N[J,J]$ and $N[\bar J, \bar J]$
implies
the vanishing of this tensor.
So imposing both constraints eqs. (\ref{con1}) and (\ref{con2}) yields
an additional integrability condition: the two complex structures
have to commute!  In the
previous section we showed that a second supersymmetry
required the existence of two complex structures, one for each chirality.
Furthermore we saw that off-shell closure of the algebra entails the two
complex structures to commute, which implies that only in this case no
further auxiliary fields are needed.  Here we get
the same result from a purely kinematic point of view:
imposing constraints of both chiralities on a general $(2,2)$ superfield
reduces the degrees of freedom of that field to those of a $(1,1)$ superfield
but integrability of these constraints requires the existence of
two, mutually commuting complex structures.

Take now $J$ and $\bar J$ two commuting complex structures.
This is sufficient to obtain full integrability
of both $J$ and $\bar J$ {\it simultanously}.  Then we can always
find a
coordinate transformation such that both $J$ and $\bar J$ are diagonal.
As the eigenvalues, $\pm i$, of $J$ and
$\bar J$ can be combined in four different ways, we get the four basic
superfields:
\begin{enumerate}
\item chiral superfield:
\begin{eqnarray}
\wh{D}\Phi\, =\, -D\Phi,\qquad\quad \wh{\bar D}\Phi\,=\, -\bar D\Phi.
\label{cc1}
\end{eqnarray}
\item anti-chiral superfield:
\begin{eqnarray}
\wh{D}\Phi\, =\, +D\Phi,\qquad\quad \wh{\bar D}\Phi\,=\, +\bar D\Phi.
\end{eqnarray}
\item twisted chiral superfield:
\begin{eqnarray}
\wh{D}\Phi\, =\, -D\Phi,\qquad\quad \wh{\bar D}\Phi\,=\, +\bar D\Phi.
\end{eqnarray}
\item twisted anti-chiral superfield:
\begin{eqnarray}
\wh{D}\Phi\, =\, +D\Phi,\qquad\quad \wh{\bar D}\Phi\,=\, -\bar
D\Phi.\label{basdef}
\end{eqnarray}
\end{enumerate}
{}From the previous, it follows that $\ker(J-\bar J) $ corresponds to
(anti-)chiral superfields and $\ker(J+\bar J)$ to twisted (anti-)chiral
superfields.
So we arrive at the main conclusion of this section:
{\it constraining both
chiralities of a general $(2,2)$ superfield reduces the degrees of freedom
to those of a $(1,1)$ superfield and there is always a coordinate
transformation such that these fields reduce to (anti-)chiral and twisted
(anti-)chiral fields}. In particular, there is no way to mimick the
spectral flow at  the level of fields, {\it i.e.} to continously
interpolate between chiral and twisted chiral fields.

Consider now a real potential $K(\phi )$ which is a function of $m$ chiral
fields
$\phi^{\alpha}$, $m$ anti-chiral fields $\phi^{\bar \alpha }$, $n$ twisted
fields $\phi^{\mu }$ and $n$ twisted anti-chiral fields $\phi^{\bar \mu}$,
$\alpha ,\bar\alpha \in \{1, \cdots, m\}$, $\mu , \bar \mu \in\{1, \cdots,
n\}$. Using eq. (\ref{basdef}), we immediately obtain the action in $(1,1)$
superspace:
\begin{eqnarray}
{\cal S} &=&\int\,d^2z d^2\theta d^2{\hat{\theta}}\, K(\phi)\nonumber\\
&=&-2 \int\,d^2zd^2\theta \left(K_{\alpha \bar\beta}(D\phi^\alpha \bar D
\phi^{\bar \beta} + D \phi^{\bar\beta}\bar D \phi^\alpha )-K_{\mu \bar\nu}
(D\phi^\mu  \bar D
\phi^{\bar \nu} + D \phi^{\bar\nu}\bar D \phi^\mu  ) \right)+\nonumber\\
&&+2\int\,d^2zd^2\theta \left(K_{\alpha \bar\nu}(D\phi^\alpha \bar D
\phi^{\bar \nu} - D \phi^{\bar\nu}\bar D \phi^\alpha )-K_{\mu \bar\beta}
(D\phi^\mu  \bar D
\phi^{\bar \beta} - D \phi^{\bar\beta}\bar D \phi^\mu  )
\right),\label{genk}
\nonumber\\
\end{eqnarray}
where
\begin{eqnarray}
K_{ab}\equiv\frac{\partial^2K(\phi)}{\partial \phi^a\partial\phi^b},
\end{eqnarray}
and $\alpha ,\bar\alpha,\beta,\bar\beta \in \{1, \cdots, m\}$,
$\mu , \bar \mu,\nu,\bar\nu \in\{1, \cdots,
n\}$. Comparing this with eq. (\ref{staction}), we can read off the metric
$g_{ab}$ and the torsion potential $b_{ab}$. Restricting ourselves to the
case where either the chiral or the twisted chiral fields are absent
yields the standard K\"ahler geometry. When both types of fields are
present we recognize from the fact that only $g_{\alpha \bar\beta}$ and
$g_{\mu \bar\nu}$ are non-vanishing the product
structure\footnote{However, this doesn't imply that the resulting manifold
can be written as the product of two manifolds. This would imply that {\it
e.g.} $g_{\alpha \bar\beta,\mu }=g_{\alpha \bar\beta,\bar\mu }=
g_{\mu\bar\nu,\alpha  }=g_{\mu \bar\nu,\bar\alpha }=0$ which is not
necessarily true here.}. Note that the
potential $K(\phi)$ in the $(2,2)$ action eq. (\ref{genk})  is only
determined modulo a generalized K\"ahler transformation:
\begin{eqnarray}
K(\phi)\simeq K(\phi)+f(\phi^{\alpha} ,\phi^{\nu}) + g(\phi^{\alpha}
,\phi^{\bar\nu}) +
\bar f(\phi^{\bar\alpha} ,\phi^{\bar\nu}) + \bar g
(\phi^{\bar\alpha} ,\phi^{\nu}).
\end{eqnarray}

Having exhausted the case where both chiralities of a general $(2,2)$
superfield were constrained, we turn in the next section to fields where
only one of the chiralities gets constrained.
\subsection{Semi-chiral superfields}
\noindent
Having a set of general superfields $\phi^a$, $a\in\{1,\cdots ,2d\}$, we
constrain only one chirality as in eq. (\ref{con1}). We perform
a coordinate transformation such that $J$ becomes diagonal:
$J^\alpha{}_{\beta}=i\delta^\alpha _\beta$ and $J^{\bar\alpha
}{}_{\bar\beta}=-i\delta^{\bar\alpha }_{\bar\beta}$. Thus we get from eq.
(\ref{con1}):
\begin{eqnarray}
\wh D \phi^\alpha =-D\phi^\alpha ,\qquad \wh{ D}\phi^{\bar\alpha }=
D\phi^{\bar\alpha }.
\end{eqnarray}
Introduce the notation
\begin{eqnarray}
\chi ^\alpha \equiv \wh{\bar D}\phi^\alpha ,\qquad
\chi ^{\bar\alpha} \equiv \wh{\bar D}\phi^{\bar\alpha},
\end{eqnarray}
with $\alpha ,\bar\alpha \in\{1,\cdots, d\}$.
We take a real potential $K(\phi)$ and we pass from $(2,2)$ to $(1,1)$
superspace,
\begin{eqnarray}
{\cal S} &=&\int\,d^2z d^2\theta d^2{\hat{\theta}}\, K(\phi)\nonumber\\
&=&-2 \int\,d^2zd^2\theta K_{\alpha \bar\beta }\left(\chi ^\alpha
D\phi^{\bar\beta} -\chi ^{\bar\beta}D\phi^\alpha \right).
\end{eqnarray}
We see that the $\chi $ fields appear algebraically, but eliminating them
through their equations of motion does not lead to a non-linear
$\sigma$-model. So we take a set of fields $\phi^a$, $a\in \{1,\cdots d\}$ and
another set $\phi^{a'}$, $a'\in \{d+1,\cdots d+ d'\}$ on which we impose the
constraints:
\begin{eqnarray}
\wh D \phi^a = iJ^a{}_BD\phi^B,\qquad \wh{\bar D}
\phi^{a'}=iJ^{a'}{}_B\bar D \phi^{B},
\end{eqnarray}
where $B\in\{1,\cdots d+d'\}$. Integrability of these constraints gives the
conditions:
\begin{eqnarray}
J^a{}_{b'}=\bar J^{a'}{}_b=0,\qquad J^a{}_{b,c'}=\bar J^{a'}{}_{b',c}=0,
\end{eqnarray}
and $J^a{}_b$ and $\bar J^{a'}{}_{b'}$ are complex structures. Through a
coordinate transformation we can diagonalize these structures. A similar
reasoning as at the beginning of this section gives that only when $d=d'$
do we get a non-linear $\sigma$-model. Furthermore the fact that both $J$
and $\bar J$ are complex structures requires $d$ to be even. So one
semi-chiral multiplet corresponds with four real dimensions.
Bringing all this together leads to the second main result of this paper:
{\it the most general non-linear $\sigma$-model in standard $N=(2,2)$
superspace is
formulated in terms of chiral, twisted chiral and semi-chiral
superfields.}

We now study such a model. We take $m$ chiral, $\phi^\mu  $ and
anti-chiral, $\phi^{\bar\mu  }$, $n$ twisted chiral $\phi^{\wt \mu }$
and twisted anti-chiral $\phi^{\wt {\bar\mu }}$ superfields. In addition
we take $d$ semi-chiral multiplets $\{ \phi^\alpha , \phi^{\bar\alpha },
\phi^{\wt\alpha }, \phi^{\wt{\bar\alpha }}\}$, with  $\mu ,\bar\mu \in
\{1,\cdots, m\}$, $\wt\mu ,\wt{\bar\mu} \in
\{1,\cdots, n\}$,   $\alpha, \bar\alpha ,\wt\alpha ,\wt{\bar\alpha } \in
\{1,\cdots, d\}$.
The defining relations
of the (anti-)chiral and twisted (anti-)chiral fields are given in eq.
(\ref{cc1}-\ref{basdef}) and the semi-chiral fields are defined by:
\begin{eqnarray}
\widehat{D}\f^\a=-D\f^\a\quad &\& &\quad
\psi^\a\equiv\widehat{\bar D}\f^\a\nonu
\widehat{D}\f^{\bar \a}=D\f^{\bar
\a}\quad &\& &\quad\psi^{\bar \a} \equiv\widehat{\bar D}\f^{\bar \a}\nonu
\chi^{\wt{ \a}}\equiv \widehat{D}\f^{\wt{\a}} \quad &\& &\quad
\widehat{\bar D}\f^{\wt{ \a}}=\bar D \f^{\wt{
\a}}\nonu
\chi^{\wt{\bar \a}}\equiv\widehat{D}\f^{\wt{\bar
\a}}\quad &\& &\quad \widehat{\bar D}\f^{\wt{\bar
\a}}=-\bar D \f^{\wt{\bar \a}}.\label{defsc}
\end{eqnarray}
Taking an arbitrary potential $K$ which is a function of all the previously
mentioned fields, we get the $\sigma$-model in $(1,1)$ superspace:
\begin{eqnarray}
{\cal S} &=&\int\,d^2z d^2\theta d^2{\hat{\theta}}\, K(\phi)\nonumber\\
&=&-2 \int\,d^2zd^2\theta\Big\{ D\Phi_cN_1\bar D\Phi_c-D\Phi_tN_2\bar D
\Phi_t- D\Phi_c N_3\bar D \Phi_t+ D\Phi_t N_3^T\bar D \Phi_c
\Big\}-\nonumber\\
&&\int\,d^2zd^2\theta\Big\{ \chi^T L\psi-\bar
D\h^TPLPD\f-\chi^T(L\psi-PL\bar  D\f-2P\wt M \bar D \h+\nonumber\\
&&2R_3\bar D\Phi_c+2R_4\bar D\Phi_t) -
(\chi^T L+D\h^TLP-2D\f^T P M+2 D\Phi^T_cR_1^TP+\nonumber\\
&& 2 D\Phi^T_tR_2^TP )\psi+2D\phi^TR_1\bar D\Phi_c-2D\phi^TR_2\bar D\Phi_t
+2 D \Phi_c^TPS_1^TP\bar D\eta +\nonumber\\
&& D\Phi_t^TPS_2^TP\bar D \eta+D\eta^TS_1\bar D \Phi_c-D \eta^TS_2\bar
D\Phi_t\Big\},
\end{eqnarray}
where $N_1$, $N_2$ and $N_3$ are $2m\times 2m$, $2n\times 2n$ and $2m\times
2n$
matrices,
\begin{eqnarray}
N_1\equiv \left( \begin{array}{cc} 0 & K_{\mu \bar\nu}\\
K_{\bar\mu \nu} & 0\end{array}\right), \qquad
N_2\equiv \left( \begin{array}{cc} 0 & K_{\wt\mu \wt{\bar\nu}}\\
K_{\wt{\bar\mu}\wt \nu} & 0\end{array}\right),\qquad
N_3\equiv \left( \begin{array}{cc} 0 & K_{\mu \wt{\bar\nu}}\\
K_{\bar\mu \wt\nu} & 0\end{array}\right),
\end{eqnarray}
where $L$, $M$ and $\wt M$ are $2d\times 2d$ matrices
\bea
L\equiv \left( \begin{array}{cc} K_{\wt\a\b} & K_{\wt\a\bar\b}\\
K_{\wt{\bar\a}\b} & K_{\wt{\bar\a}\bar\b}\end{array}\right), \qquad
\wt M\equiv \left( \begin{array}{cc} 0 & K_{\wt\a\wt{\bar\b}}\\
K_{\wt{\bar\a}\wt\b} & 0\end{array}\right),\qquad
M\equiv \left( \begin{array}{cc} 0 & K_{\a{\bar\b}}\\
K_{{\bar\a}\b} & 0\end{array}\right),\label{defmat}
\eea
$S_1$, a $2d\times 2m$ matrix, $S_2$ a $2d\times 2n$ matrix,
\bea
S_1\equiv \left( \begin{array}{cc} K_{\wt\a\mu } & K_{\wt\a\bar\mu }\\
K_{\wt{\bar\a}\mu } & K_{\wt{\bar\a}\bar\mu}\end{array}\right), \qquad
S_2\equiv \left( \begin{array}{cc} K_{\wt\a\wt\mu } & K_{\wt\a\wt{\bar\mu} }\\
K_{\wt{\bar\a}\wt\mu } & K_{\wt{\bar\a}\wt{\bar\mu}}\end{array}\right),
\label{defmat1}
\eea
and the $2d\times 2m $ matrices $R_1$ and $R_3$ and the
$2d\times 2n $ matrices $R_2$ and $R_4$  together with $P$:
\bea
&&R_1\equiv \left( \begin{array}{cc} 0 & K_{\alpha \bar\mu }\\
K_{\bar\alpha \mu } & 0\end{array}\right), \qquad
R_2\equiv \left( \begin{array}{cc} 0 & K_{\alpha \wt{\bar\mu} }\\
K_{\bar\alpha\wt \mu } & 0\end{array}\right), \qquad
R_3\equiv \left( \begin{array}{cc} -K_{\wt\alpha \mu } & 0\\
0 & K_{\wt{\bar\alpha }\bar\mu} \end{array}\right),\nonumber\\
&&R_4\equiv \left( \begin{array}{cc} 0 &- K_{\wt\a\wt{\bar\mu}}\\
K_{\wt{\bar\a}\wt\mu} &0\end{array}\right),\qquad
P\equiv \left( \begin{array}{cc}{\bf 1} & 0\\
0 & -{\bf 1}\end{array}\right) .
\label{defmat2}
\eea
$\Phi_c$ and $\Phi_t$ are $2m\times 1$ and $2n\times 1$ matrices resp.
and $\f$ and $\h$ are $2d\times 1$ matrices:
\bea
\Phi_c\equiv \left( \begin{array}{c} \phi^\mu  \\ \phi^{\bar\mu
}\end{array}\right),
\qquad
\Phi_t\equiv \left( \begin{array}{c} \phi^{\wt\mu} \\ \phi^{\wt{\bar\mu}}
\end{array}\right),\qquad
\f\equiv\left( \begin{array}{c} \f^\a \\ \f^{\bar \a}\end{array}\right),
\qquad
\h\equiv\left( \begin{array}{c} \f^{\wt \a} \\ \f^{\wt{\bar
\a}}\end{array}\right).
\eea
Also in this case we get that the potential is determined modulo
a generalized K\"ahler transformation:
\begin{eqnarray}
K(\phi)\simeq K(\phi)+f(\phi^{\mu } ,\phi^{\wt\mu},\phi^\alpha ) +
g(\phi^{\mu} ,\phi^{\wt{\bar\mu}},\phi^{\wt{\bar\alpha }}) +
\bar f(\phi^{\bar\mu } ,\phi^{\wt{\bar\mu}},\phi^{\bar\alpha }) + \bar g
(\phi^{\bar\mu } ,\phi^{\wt\mu},\phi^{\wt\alpha }). \label{scgk}
\end{eqnarray}
One finds that the equations of motion for $\chi$ and $\psi$ are
\bea
L\psi&=&PL\bar D \f+ 2 P \wt M \bar D \h-
2R_3\bar D\Phi_c-2R_4\bar D\Phi_t,\nonu
\chi^TL&=&-D\h^TLP+2D\f^T P M
-2 D\Phi^T_cR_1^TP-
2 D\Phi^T_tR_2^TP .\label{eom11}
\eea
Assuming that $L$ is invertible, one
solves  eq. (\ref{eom11}) for the auxiliary fields $\psi$ and $\chi$. The
action in the
second order formalism is given by
\begin{eqnarray}
{\cal S} &=&-2 \int\,d^2zd^2\theta\Big\{ D\Phi_cN_1\bar D\Phi_c-D\Phi_tN_2\bar
D
\Phi_t- D\Phi_c N_3\bar D \Phi_t+ D\Phi_t N_3^T\bar D \Phi_c
\Big\}+\nonumber\\
&&\int\,d^2zd^2\theta\Big\{ \bar
D\h^TPLPD\f -2D\phi^TR_1\bar D\Phi_c+2D\phi^TR_2\bar D\Phi_t
-2 D \Phi_c^TPS_1^TP\bar D\eta -\nonumber\\
&& D\Phi_t^TPS_2^TP\bar D \eta-D\eta^TS_1\bar D \Phi_c+ D \eta^TS_2\bar
D\Phi_t +
\left(D\h^TLP-2D\f^T P M+\right. \nonumber\\
&&\left. 2 D\Phi^T_cR_1^TP+
2 D\Phi^T_tR_2^TP\right) L^{-1} \left( PL\bar D \f+ 2 P \wt M \bar D \h-
2R_3\bar D\Phi_c-2R_4\bar D\Phi_t\right)\Big\}.\nonumber\\
\label{ac1o}
\end{eqnarray}
Introducing the notation
\bea
\F\equiv \left(\begin{array}{c}\f\\ \h\\ \Phi_c\\ \Phi_t \end{array}\right),
\eea
one reads off the complex structures:
\bea
J&=&\left(\begin{array}{cccc}iP&0&0&0\\
-2iL^{-1}{}^T M P & i L^{-1}{}^T
P  L^T &2iL^{-1}{}^TPR_1 &2iL^{-1}{}^TPR_2 \\
0&0&iP&0\\
0&0&0&iP
\end{array}\right)\nonu
\bar J &=&\left(\begin{array}{cccc}-iL^{-1}PL& -2iL^{-1} P \wt M&
2iL^{-1}R_3 &2i L^{-1}R_4 \\
0&-iP& 0&0  \\
0&0&iP&0\\
0&0&0&-iP \end{array}\right).\label{cs1}
\eea
\setcounter{equation}{0}
\section{$N=(2,2)$ non-linear $\sigma$-models}
An obvious question which now arises is whether all $N=(2,2)$ non-linear
$\sigma$-models can be described using the previously constructed superfields.
An important result obtained in
\cite{martinnew} states that $\ker (J-\bar J)$ and $\ker (J+\bar J)$ are
always
integrable. This allows us to parametrize these kernels by chiral and
twisted-chiral fields resp. If we now choose such preferred coordinates we
can restate the result by saying that the levels where these coordinates
are constant yield submanifolds where $(J\pm\bar J)$
are non-degenerate. So whatever is left should then be described by
semi-chiral fields.

For simplicity, we focus on manifolds where
$\ker [J,\bar J]=\emptyset$. One would expect those to be described by
semi-chiral superfields. Let us first determine what the implications are
of a $\sigma$-model solely described by semi-chiral fields and where
$\ker [J,\bar J]\neq\emptyset$. Given an arbitrary vector $\xi$,
we decompose it as $\xi=\xi_0+\xi_1$ where $[J,\bar J]\xi_0=0$ and
$[J,\bar J]\xi_1\neq 0$.
Writing the $\xi_0^a$ components as $\zeta$ and the $\xi_0^{\tilde a}$
components as $\tilde\zeta$, one derives from eq. (\ref{cs1})
that $\xi_0$ satisies either
\begin{eqnarray}
\{P,L\}\zeta =-2P\wt M\tilde\zeta ,\qquad \{P,L^T\}\tilde \zeta = -2 P
M\zeta ,
\end{eqnarray}
or
\begin{eqnarray}
[P,L]\zeta =-2P\wt M\tilde\zeta ,\qquad [P,L^T]\tilde \zeta = -2 P
M\zeta .
\end{eqnarray}
Using this and the explicit form for the action, eq. (\ref{ac1o}),
one shows that the metric is
degenerate: $g(\xi_0,\xi')=0$, where $\xi'$ is an arbitrary vector.
Presumably, having $\ker [J,\bar J]\neq\emptyset$ points towards the existence
of gauge invariances such as were studied in \cite{gauge}.

{}From now on we
assume that
the metric is non-degenerate, and as such that $\ker [J,\bar J]=\emptyset$. A
necessary
and sufficient condition for the latter is
\begin{eqnarray}
\det {\cal N}_1\neq 0,\qquad  \det {\cal N}_2\neq 0,
\end{eqnarray}
with
\begin{eqnarray}
{\cal N}_1\equiv \left( \begin{array}{cc} K_{\wt\alpha \beta} &
K_{\wt\alpha \wt{ \bar\beta}}\\
K_{\bar\alpha \beta} & K_{\bar\alpha \wt{ \bar\beta}}\end{array}\right),
\qquad
{\cal N}_2\equiv \left( \begin{array}{cc} K_{\wt\alpha \bar\beta} &
K_{\wt\alpha  \wt{\bar\beta}}\\
K_{\alpha \bar\beta} & K_{\alpha \wt{\bar\beta}}\end{array}\right).
\end{eqnarray}

We turn back to a complex manifold where  $\ker [J,\bar J]=\emptyset$.
A necessary condition for a description of such a manifold in terms of
semi-chiral fields is that its (real) dimension is a multiple of
4. Let us give a local argument why this is true. Consider the
following problem: given a $2n\times 2n $ matrix $J_+$, such
that
$J_+^2= -{\bf 1}$.
We choose the canonical form for it:
\begin{eqnarray}
J_+=i\left(\begin{array}{cc}{\bf 1}&0\\0&-{\bf 1}
\end{array}\right).
\end{eqnarray}
Consider now a second  $2n\times 2n $ matrix $J_-$, such that
the standard (flat) metric is hermitean. Its most general form is given by
\begin{eqnarray}
J_-=\left(\begin{array}{cc}A&B\\B^*&A^*
\end{array}\right),\qquad\mbox{with }A^\dagger =-A,\quad B^T=-B.
\end{eqnarray}
We define ${\cal C}$ as
\begin{eqnarray}
{\cal C}\equiv [J_+,J_-]=\left(\begin{array}{cc}0&2iB\\-2iB^*&0
\end{array}\right), \label{fgr}
\end{eqnarray}
and assume that $\ker {\cal C}=\emptyset$, or in other words $\det {\cal C}
\neq 0$. One computes,
\begin{eqnarray}
\det {\cal C}=(-1)^n 4^n |\det B|^2. \label{det}
\end{eqnarray}
As ${\cal C}$ is anti-hermitean, it has imaginary eigenvalues. Furhermore
from the form of ${\cal C}$, eq. (\ref{fgr}), one easily gets that if
$\lambda$ is an eigenvalue, then so is $-\lambda$. Combining all of this
yields $\det {\cal C}>0$.  Comparing this to eq. (\ref{det}), requires
$n$ to be even.

Having established that the dimension of our manifold is a multiple of 4,
we will now put $J$ in its standard form
\begin{eqnarray}
J=i\left(\begin{array}{cc}{\bf 1}&0\\0&-{\bf 1}
\end{array}\right).
\end{eqnarray}
The constancy of $J$ implies
\begin{eqnarray}
\Gamma^\alpha _-{}_{c\bar\beta }&=&0,\nonumber\\
\Gamma_-{}_{\alpha \beta\bar\gamma}&=&-2 T_{\alpha\beta\bar\gamma},
\nonumber\\
\Gamma_-{}_{\bar\alpha \beta\gamma}&=&g_{\bar\alpha \beta,\gamma},\nonumber\\
T_{\alpha\beta\gamma}&=&0. \label{conJ}
\end{eqnarray}
The vanishing of the Nijenhuis tensor for $J$ imposes no further restrictions.
{}From eq. (\ref{conJ}), one obtains that locally, the metric and the
torsion can be expressed in terms of a vector potential $k$:
\begin{eqnarray}
g_{\alpha \bar\beta}&=&\frac 1 2 (k_{\alpha ,\bar\beta}+k_{\bar\beta,\alpha
}),  \nonumber\\
T_{\alpha \beta\bar\gamma}&=&-\frac 1 4 (k_{\alpha ,\beta}-
k_{\beta,\alpha })_{,\bar\gamma}, \label{defk}
\end{eqnarray}
and the complex conjugates of these expressions. The vectorfield $k_a$
is determined
modulo
\begin{eqnarray}
k_\alpha \simeq k_\alpha + f_\alpha +ig_{,\alpha },
\end{eqnarray}
where $f_{\alpha ,\bar\beta}=0$ and $g$ is a real scalar function.

We now analyze the consequences for $\bar J$.
As $[J,\bar J]$ is non-degenerate, we can parametrize $\bar J$ as
\begin{eqnarray}
\bar J=\left(\begin{array}{cc}a&b\\-b^{-1}(1+a^2)&-b^{-1}ab
\end{array}\right),\label{cs2o}
\end{eqnarray}
and $a^2\neq -{\bf 1}$.
Hermiticity of the metric is satisfied iff.
\begin{eqnarray}
b_{\alpha \beta}=-b_{\beta\alpha },\qquad a^\alpha
{}_\gamma b^{\gamma\beta}+ a^\beta{}_\gamma b^{\gamma\alpha
}=0.\label{uuu}
\end{eqnarray}
The constancy of $\bar J$ gets translated into the following conditions:
\begin{eqnarray}
&&b^{\alpha \beta}{}_{,\bar \gamma}=0.\nonumber\\
&&T_{\alpha \beta\bar\gamma}=-\frac 1 2 (b^{-1}a)_{\alpha
\beta,\bar\gamma},\label{bjcon}
\end{eqnarray}
and their complex conjugate.
Finally, the vanishing of the Nijenhuis tensor imposes further restrictions.
Introducing the notation
\begin{eqnarray}
x_{\alpha \beta}\equiv \left( b^{-1}(a+i)\right)_{\alpha \beta},\quad
x_{\bar\alpha \bar\beta}\equiv x^*_{\alpha \beta}= \left((a-i)^{-1}b
\right)_{\bar\alpha \bar\beta},
\end{eqnarray}
one concisely writes the Nijenhuis conditions as
\begin{eqnarray}
x^{\bar\delta}{}_{[\alpha }x_{\beta\gamma],\bar\delta}=
x_{\bar\delta\bar\varepsilon,[\alpha }
x^{\bar\delta}{}_\beta x^{\bar\varepsilon}{}_{\gamma]}.\label{bjnij}
\end{eqnarray}

We now want to compare this with what we get from a semi-chiral description.
The easiest way to obtain the vectors $k$ is to pass to $(2,1)$ superspace.
In general, if one has a $N=(2,1)$ non-linear
$\sigma$-model, one gets by combining eqs.
(\ref{staction}) and (\ref{defk}) that it is given in $N=(2,1)$ superspace by
\begin{eqnarray}
{\cal S}=\frac 1 2 \int d^2z d^2\theta d\hat\theta\left(
k_\alpha \bar D \phi^\alpha -k_{\bar\alpha }\bar D \phi^{\bar\alpha }
\right),
\end{eqnarray}
where $\phi$ are $(2,1)$ chiral fields:
\begin{eqnarray}
\widehat{D}\phi^\alpha =-D\phi^\alpha ,\qquad
\widehat{D}\phi^{\bar\alpha} =D\phi^{\bar\alpha}.
\end{eqnarray}
Starting now from a potential in terms of semi-chiral fields, we perform the
integration over $\hat{\bar\theta}$ and obtain
\begin{eqnarray}
{\cal S}=\int d^2z d^2\theta d\hat\theta\left(
K_a\psi^a+K_{\wt{\alpha }}\bar D \phi^{\wt{\alpha }}-
K_{\wt{\bar\alpha }}\bar D \phi^{\wt{\bar\alpha }}
\right).\label{ac21}
\end{eqnarray}
The equations of motion for $\psi$ imply that $\hat\phi_a\equiv K_a$ is a
chiral
(2,1)
field. This is nothing but a reflection of the fact that
the coordinate transformation
\begin{eqnarray}
\phi^{\alpha}\rightarrow\hat{\phi}^{\alpha}=\phi^{\alpha}&&
\phi^{\tilde{\alpha}}\rightarrow\hat{\phi}_{{\alpha}}=
K_{\alpha}\nonumber\\
\phi^{\bar\alpha}\rightarrow\hat{\phi}^{\bar\alpha}= \phi^{\bar\alpha}
&&\phi^{\tilde{\bar\alpha}}\rightarrow\hat{\phi}_{{\bar\alpha}}=
K_{\bar\alpha},
\end{eqnarray}
diagonalizes $J$. Obviously this coordinate transformation is not
compatible with the original semi-chiral nature of the fields. Integrating
over $\psi$ in eq. (\ref{ac21}), we get the (2,1) action:
\begin{eqnarray}
{\cal S}=\int d^2z d^2\theta d\hat\theta\left(
K_{\wt{a}}P^{\wt{a}}{}_{\wt{b}}(K^{-1})^{\wt b c}(\bar D \hat\phi_c -
K_{cd}\bar
D\hat\phi^d)
\right),
\end{eqnarray}
from which one reads the vectors $k$. In terms of the {\it original}
semi-chiral
coordinates, they become particularly simple
\begin{eqnarray}
{\wt k} =LPL^{-1}P{\wt K},\qquad k=2ML^{-1}{\wt k}, \label{kcond}
\end{eqnarray}
where
\begin{eqnarray}
{\wt K}\equiv\left(\begin{array}{c}K_{\wt\alpha}\\K_{\wt{\bar\alpha}}
\end{array}\right),\quad
{\wt k}\equiv\left(\begin{array}{c}k_{\wt\alpha}\\k_{\wt{\bar\alpha}}
\end{array}\right),\quad
{k}\equiv\left(\begin{array}{c}k_{\alpha}\\k_{{\bar\alpha}}
\end{array}\right).
\end{eqnarray}
So in order to show that a description in terms of semi-chiral fields is
possible, one has to show that there exists a coordinate system such that
the solution to eqs. (\ref{uuu}), (\ref{bjcon}) and (\ref{bjnij}) is given
by eq. (\ref{kcond}).

Though we are not able to show this, we focus on the special case where
the Nijenhuis conditions are trivially satisfied. This requires that the
torsion
vanishes, which, as eq.
(\ref{bjcon}) shows, is so if
$a$ in eq. (\ref{cs2o}) vanishes. Combining this with
$\{J,\bar J\}=0$ and eq. (\ref{kahler})
gives that the manifold is hyper-K\"ahler. In fact for any hyper-K\"ahler
manifold, one has $\ker [J,\bar J]= \emptyset$. Note however that we do
know how to put $\sigma$-models on hyper-K\"ahler manifolds in $(2,2)$
superspace, after all they are K\"ahler manifolds and can be described in
terms of chiral fields. However this is only possible if we choose the
left complex structure to be the same as the right one. It is
clear that on a hyper-K\"ahler manifold, the left and right complex
structure can be chosen diferently. Precisely this case is being
investigated here. Using eq. (\ref{cs1}), one shows that the necessary and
sufficient conditions for $J$ and $\bar J$ to anti-commute are given by:
\begin{eqnarray}
&&L^{-1}PLP+PL^{-1}PL=4L^{-1}P\wt ML^{-1T}MP\nonumber\\
&&\{ P, L^{-1T}MPL^{-1}\}=\{P,L^{-1}P\wt M L^{-1T}\}=0.
\end{eqnarray}
Restricting ourselves to $d=4$, we find that the two latter eqs. are
trivially satisfied while the former becomes:
\begin{eqnarray}
|K_{\phi\eta}|^2+|K_{\phi\bar \eta}|^2=2K_{\eta\bar \eta}K_{\phi\bar\phi}.
\label{hypkah}
\end{eqnarray}
It is
known that a 4-dimensional K\"ahler manifold is hyper-K\"ahler iff. the
K\"ahler potential satisfies the Monge-Amp\`ere equation. So a concrete
way to test our hypothesis would be to show that there is a coordinate
system where eq. (\ref{hypkah}) is equivalent to the Monge-Amp\`ere
equation. In \cite{icproc}, we checked explicitely that a non-trivial class
of hyper-K\"ahler manifolds, the special hyper-K\"ahler
manifolds \cite{stef}, allows for a semi-chiral parametrization.
If it turns out to be true that all hyper-K\"ahler manifolds can be
described with semi-chiral coordinates, then this would provide a
non-trivial result: not only the metric can be computed from the
semi-chiral potential but all three complex structures as well!

To end this section, we answer the question whether for hyper-K\"ahler
manifolds, the K\"ahler and the semi-chiral potential are identical. An
easy manifold to verify this is flat $4n$-dimensional space. The K\"ahler
potential is given by
\begin{eqnarray}
K_{Kahler}=\sum_{i=1}^n(x^i\bar x^i + v^i\bar v^i),
\end{eqnarray}
and labeling rows and columns as $x,$ $\bar x$, $v$, $\bar v$ we have the
complex
structures:
\begin{eqnarray}
J=\left(\begin{array}{cc}i\sigma_3&0\\0&i\sigma_3
\end{array}\right),\quad \bar J= \left(\begin{array}{cc}0&-\sigma_2\\
\sigma_2&0
\end{array}\right) .
\end{eqnarray}
A coordinate transformation which brings us to the semi-chiral
parametrization is
\begin{eqnarray}
x^i\rightarrow\phi^i=x^i,&&  \bar x^i\rightarrow\bar\phi^i=\bar
x^i\nonumber\\
v^i\rightarrow \eta^i=\bar x^i+v^i , &&\bar v^i\rightarrow \bar
\eta^i= x^i+\bar v^i,
\end{eqnarray}
and labeling rows and columns as $\phi$, $\bar\phi$, $\eta$ and $\bar\eta$
we get the complex structures in semi-chiral coordinates:
\begin{eqnarray}
J=\left(\begin{array}{cc}i\sigma_3&0\\2\sigma_2&i\sigma_3
\end{array}\right),\quad \bar J= \left(\begin{array}{cc}-i\sigma_3&-\sigma_2\\
0&-i\sigma_3
\end{array}\right) .
\end{eqnarray}
Comparing the complex structures and metric to the general expressions
yields the potential:
\begin{eqnarray}
K_{semi}\propto\sum_{i=1}^n
(2\phi^i\bar\phi^i+\eta^i\bar\eta^i-2\phi^i\eta^i-2\bar\phi^i\bar\eta^i).
\end{eqnarray}
Taking into account that the K\"ahler potential is only defined modulo a
K\"ahler transformation and the semi-chiral potential has an ambiguity
expressed in eq. (\ref{scgk}), one verifies that both potential are {\it not}
equivalent.

\setcounter{equation}{0}
\section{Examples}
Supersymmetric Wess-Zumino-Witten models on even dimensional groupmanifolds
are interesting examples of $(2,2)$ supersymmetric non-linear
$\sigma$-models. They are conformally invariant, they have a large number
of isometries and they are torsionful.
We can write $J=L^{-1}jL$ and $\bar J= R^{-1}\bar j R$ where
$L$ and $R$ are the left, right resp., invariant vielbeins. The (constant)
complex
structures $j$ and $\bar j$ act on the Lie algebra.

In \cite{usold}, the
integrability conditions for a complex structure on a semi-simple Lie
algebra were solved. The action of the complex structure is almost
completely determined by a Cartan decomposition: the complex structure has
eigenvalue $+i$ ($-i$ resp.) on generators corresponding with positive
(negative resp.) roots. The only freedom left is the action of the complex
structure on the Cartan subalgebra. Except for the requirement that the
structure maps the CSA bijectively to itself, no further conditions have to be
imposed. As any two Cartan decompositions are related through an inner
automorphism, we can state that the complex structure on a Lie algebra is
uniquely determined except for its action on the CSA which is left
invariant by the complex structure.

However the resulting left and right complex structures on the {\it group}
do not necessarily commute. In fact, in \cite{RSS} it was shown that only
on $SU(2)\times U(1)$ a choice for $j$ and $\bar j$ can be made such that
$[J,\bar J]=0$. There a description in terms of a chiral and a twisted
chiral field is possible.

In the next we give 3 examples of $(2,2)$ WZW models in superspace: the
$ SU(2) \times U(1) $ model in terms of a chiral and a twisted chiral
field, making a different choice for the complex structures we obtain
the same model but now in terms of a semi-chiral multiplet and
finally $SU(2)\times SU(2)$ as an example of a model having both
a chiral and a semi-chiral multiplet.
\subsection{The $ SU(2) \times U(1) $ WZW model in terms of a chiral
and a twisted chiral multiplet}
\noindent We give a brief summary of the results of \cite{RSS}. There it was
shown that
for a particular choice of the complex structures one could formulate the
model in terms of a chiral $\phi$, an anti-chiral $\bar\phi$, a twisted
chiral $\chi $ and a twisted anti-chiral $\bar\chi $ superfield. A group
element is written as
\begin{eqnarray}
g=\frac{e^{i\theta}}{\sqrt{|\phi|^2+|\chi |^2}}  \left( \begin{array}{cc}
                         \chi  &\bar\phi \\
                         -\phi & \bar\chi
                         \end{array} \right),
\end{eqnarray}
with
\begin{eqnarray}
\theta=-\frac 1 2 \ln \left(|\phi|^2+|\chi |^2\right).
\end{eqnarray}
The potential is then given by
\begin{eqnarray}
K=-\int^{\frac{|\chi |^2}{|\phi|^2}}\frac{d\zeta }{\zeta }\ln (1+\zeta )
\, +\, \frac 1 2 \left( \ln (\phi\bar\phi)\right)^2 .
\end{eqnarray}
In order to achieve this one has to make a different choice for the left
and the right complex structures on the {\it Lie algebra}. The only
difference resides in its action on the CSA: there it differs by a sign.
\subsection{The $ SU(2) \times U(1) $ WZW model in terms of a semi-chiral
multiplet}
In \cite{martinnew} an implicit description of the $SU(2) \times U(1) $ WZW
model was given but
now in terms of a semi-chiral multiplet. Here we give the explicit
description of the model using one semi-chiral multiplet.  For this we now
choose the complex structures on the Lie algebra to be equal.

We parametrize $SU(2) \times U(1)$ by:
\be
g=e^{\frac{i}{2} \theta_L \s_3} \, e^{\frac{i}{2} \zeta \s_2} \,
   e^{\frac{i}{2} \theta_R \s_3} \, e^{\frac{i}{2} \chi \s_0},
\ee
with $\s_i \, (i=1,2,3)$ the  Pauli matrices en $\s_0$ the $2 \times 2$
unit matrix. We now introduce a semi-chiral multiplet parametrized by
$\phi$, $\bar\phi$, $\eta$ and $\bar\eta$. These fields are related to the
previously introduced coordinates by
\begin{eqnarray}
\phi=\bar z_1,\qquad \eta=z_1-\bar z_2,\qquad \bar\phi=z_1,\qquad
\bar\eta=\bar z_1-z_2,
\end{eqnarray}
where,
\begin{eqnarray}
         z_1 &=& i \chi - 2 i \ln \cos \zeta /2 + \theta_L + \theta_R,
\nonumber
\\
         z_2 &=& -i \chi + 2 i \ln \sin \zeta /2 + \theta_L - \theta_R,
\end{eqnarray}
and ${\bar z}_1$, ${\bar z}_2$ are given by the complex conjugate of these.
The inverse transformations are,
\begin{eqnarray}
         \zeta &=& 2 \arctan (\exp-\frac{i}{4} ( z_1-\bar{z}_1 +
z_2-\bar{z}_2)),\nonumber\\
         \theta_L &=&  \frac{1}{4} ( z_1 + \bar{z}_1 +  z_2 +
\bar{z}_2),\nonumber\\
         \theta_R &=&  \frac{1}{4} ( z_1 + \bar{z}_1 -  z_2 -
\bar{z}_2),\nonumber\\
         \chi &=& -\ln ( \exp  \frac{i}{2} ( z_1 - \bar{z}_1 )
                        + \exp  \frac{i}{2} ( -z_2 + \bar{z}_2 ) ).
\end{eqnarray}
Ordening the coordinates as
$(\f,\bar{\f},\eta,\bar{\eta})$, one computes the complex
structures:
\bea
J&=& i \left( \begin{array}{cccc}
                     1 & 0 & 0 & 0 \\
                     0 & -1 & 0 & 0 \\
                     0& -2  & 1 & 0 \\
                     2 & 0 & 0 & -1
                   \end{array}   \right), \\
\bar{J} &=& i \left( \begin{array}{cccc}
                     -1 & 0 & 0 & 2 \sin^2 \zeta /2 \\
                     0 & 1 & -2 \sin^2 \zeta /2 & 0 \\
                     0 &  0 & -1   & 0 \\
                     0 & 0 & 0 & 1
                   \end{array}   \right).
\eea
The potential can be calculated and this gives the following simple
expression involving a dilogarithm,
\begin{eqnarray}
K=-\f \bar{\f} + \bar{\f} \bar{\eta} + \f \eta
-2 i \int^{\bar{\eta}-\eta} dx \, \ln ( 1+ \exp \frac{i}{2} x).
\end{eqnarray}
\subsection{The $ SU(2) \times SU(2) $ WZW model}
As a last example we consider the $ SU(2) \times SU(2) $ WZW model, which
has not been put in superspace yet.
Each $SU(2)$ factor gets parametrized using  Eulerangles, $
g_j=e^{\frac{i}{2} \theta_{Lj} \s_3} \, e^{\frac{i}{2} \phi_j \s_2} \,
e^{\frac{i}{2} \theta_{Rj} \s_3}$, with $j\in\{1,2\}$.
Again we choose the left and right complex structures  on the Lie algebra
to be equal. The final coordinates are the
semi-chiral multiplet $ ( \phi , \bar{\phi} , \eta , \bar{\eta} ) $ and
the chiral multiplet $ ( \zeta , \bar{\zeta} ) $.
We introduce auxiliary coordinates $x_i$, $i\in\{1,\cdots 6\}$, related to the
original ones by
\bea
x_1 &=& 2 \ln \sin \frac{\phi_1}{2} - \theta_{L2} + \theta_{R2}, \nonumber \\
x_2 &=& \theta_{L1},\qquad
x_3\ =\ \theta_{R1}, \nonumber \\
x_4 &=& 2 \ln \sin \frac{\phi_2}{2} - \theta_{R1} + \theta_{L1}, \nonumber \\
x_5 &=& \theta_{L2},\qquad
x_6\ =\ \theta_{R2},
\eea
with inverse given by
\bea
\phi_1 &=& 2 \arcsin \exp (\frac{1}{2} (x_1+x_5-x_6)), \nonumber \\
\phi_2 &=& 2 \arcsin \exp (\frac{1}{2} (x_4+x_3-x_2)).
\eea
The final coordinates are related to these by
\bea
\zeta &=& -i x_1+x_4, \nonumber \\
\phi &=& \frac{x_2}{2} +\frac{i}{4} \ln (1-\exp -(x_1+x_5-x_6)), \nonumber \\
\eta &=& \frac{x_6}{2} +\frac{i}{4} \ln (1-\exp -(x_3+x_4-x_2)), \nonumber \\
\eea
with $\bar{\zeta}$, $\bar{\phi}$ and  $\bar{\eta}$
given by the complex conjugate of these expressions. The inverse
transformations are
\bea
x_1 &=& \frac{i}{2}(\zeta-\bar{\zeta}),\qquad
x_2\ =\ \phi+\bar{\phi},              \nonumber \\
x_4 &=& \frac{1}{2}(\zeta+\bar{\zeta}), \qquad
x_6\ =\ \eta+\bar{\eta},             \nonumber \\
x_5 &=& -\frac{i}{2}(\zeta-\bar{\zeta}) + \eta + \bar{\eta}
        -\ln(1-\exp (2i(\bar{\phi}-\phi))),\nonumber \\
x_3 &=& -\frac{1}{2}(\zeta+\bar{\zeta}) + \phi + \bar{\phi}
        -\ln(1-\exp (2i(\bar{\eta}-\eta))).\nonumber
\eea
Ordering the coordinates as
$(\phi,\bar{\phi},\eta,\bar{\eta},\zeta,\bar{\zeta})$,
we get the complex structures in a recognizable form, where
$\alpha_1 = \frac{1}{2}  \sec^2 \phi_1 /2 $ and
$ \alpha_2 = \frac{1}{2} \sec^2 \phi_2  /2$ :
\bea
J^b{}_a &=&  \left( \begin{array}{cccccc}
                     i & 0 & 0 & 0 & 0 & 0 \\
                     0 & -i & 0 & 0 & 0 & 0\\
                     0 &  1 / \alpha_1 & i & 0 & 0 & - \frac{1}{2} \\
                     1 / \alpha_1 & 0 & 0 & -i & - \frac{1}{2}  & 0 \\
                     0 & 0 & 0 & 0 & i & 0 \\
                     0 & 0 & 0 & 0 & 0 & -i
                   \end{array}   \right), \\
\bar{J}^b{}_a &=&  \left( \begin{array}{cccccc}
                    -i & 0 & 0 & -1 / \alpha_2  &i/2   & 0 \\
                    0 & i & - 1 / \alpha_2 & 0 & 0 & - i/2 \\
                    0 & 0 & -i & 0 & 0 & 0  \\
                    0 & 0 & 0 & i & 0 & 0 \\
                    0 & 0 & 0 &  0 & i   & 0 \\
                    0 & 0 & 0 & 0 & 0 & -i
                   \end{array}   \right).
\eea
Comparing the structures and
the action with the general expressions, we get second order differential
equations which can be solved for the potential, which again gets
expressed in terms of dilogarithmic integrals:
\bea
K &=& -\zeta \bar{\zeta} + \zeta \bar{\phi} + \bar{\zeta} \phi + i \eta \zeta
   - i \bar{\eta} \bar{\zeta}  + i \bar{\eta} \bar{\phi} - i \eta \phi
   \nonumber \\
 &&  - i \int^{\bar{\phi}-\phi} dy \, \ln (1-\exp i y)
   - i \int^{\bar{\eta}-\eta} dy \, \ln (1-\exp i y) ,
\eea
where we performed a trivial rescaling of the fields by a factor 2.
\section{Conclusions}
\noindent
We showed that chiral, twisted chiral and semi-chiral superfields provide
the exhaustive list of $(2,2)$ superfields. A very interesting
question is whether this is sufficient to describe any
$N=(2,2)$ non-linear $\sigma$-model. We provided several pieces of
evidence for an affirmative answer to this question.
An important clue is the commutator
of the left and right complex structures.
One can show that $\ker [J,\bar J]=\ker (J-\bar J)\oplus
\ker (J+\bar J)$ can always be integrated to chiral and twisted chiral
superfields resp.

Remains the subspace where the commutator is
non-degenerate. This should then be integrable to semi-chiral fields. The
requirement that the metric of the resulting $\sigma$-model is
non-degenerate imposes that $\ker [J,\bar J]= \emptyset$ in the
semi-chiral directions. Furthermore one has that one semi-chiral multiplet
corresponds to four  real dimensions. We showed that the dimension
of the subspace of a complex manifold where the commutator is non-degenerate
is indeed a multiple of four. We obtained explicitely the conditions on the
complex manifold with $\ker [J,\bar J]= \emptyset$, under
which the $\sigma$-model can be described by semi-chiral superfields. It
remains to be shown that all such manifolds indeed satisfy these conditions.
We pointed out a particularly interesting and simple subcase where this
hypothesis can be tested: 4 dimensional hyper-K\"ahler manifolds, where one
makes
a different choice for the left and the right complex structures. A necessary
and sufficient condition for a 4 dimensional K\"ahler manifold to be
hyper-K\"ahler,
is that the K\"ahler potential satisfies the Monge-Amp\`ere equation. We
derived the condition on the semi-chiral potential which guarantees it to be
hyper-K\"ahler. It turns out to be similar
to the Monge-Amp\`ere equation. If hyper-K\"ahler manifolds can indeed be
described by semi-chiral coordinates, then one has a potential, different
from the K\"ahler potential, which does not only allow for the computation
of the metric but of the three fundamental two-forms as well.

If  all $\sigma$-models can indeed be described by
$(2,2)$ superfields, then {\it all} $N=(2,2)$
manifolds are locally charactarized by a scalar potential. Besides the
mathematically
very intriguing
implication that the geometry of
a large class of complex manifolds is locally determined by a single
potential,
this opens
interesting physics perspectives as well.
In particular, it would allow the systematic
study of (2,2), (2,1) and (2,0) strings. Up to now, the only $N=2$ strings
studied are those described solely by chiral fields \cite{ov} and those
described by chiral and twisted chiral fields \cite{ch}. The geometry of
$N=2$ strings with semi-chiral fields is presently being investigated.
An interesting question which arises in this context, in particular for
those manifolds, {\it e.g.} the hyper-K\"ahler ones,
which allow for different choices for the left and right complex structures,
is in how far the
geometry of an $N=2$ string depends on the choice of the complex
structures.
Such a study would be relevant for the recent proposals in \cite{emil}
relating the $D=11$ membrane to the type IIB stringtheory.

Another point which certainly deserves further attention is a systematic
study of $T$-duality, such as was done in \cite{kiritsis} for chiral and
twisted
chiral fields, which includes semi-chiral fields.

\vspace{5mm}

\noindent {\bf Acknowledgments}: We would like to thank Chris Hull, Elias
Kiritsis, Costas Kounnas, Walter Troost and  Toine Van Proeyen for several
useful discussions. We specially thank Martin Ro\v{c}ek for keeping our
interest in this problem alive and for many discussions. J.T. thanks the
Institute for Theoretical Physics at Leuven, where a significant part of this
work was performed, for its hospitality. This work was supported in part by the
European Commission TMR programme ERBFMRX-CT96-0045 in which both authors are
associated to K. U. Leuven.

\vspace{5mm}

\renewcommand{\theequation}{A.\arabic{equation}}
\setcounter{equation}{0}
\par \noindent
  {\bf A. Appendix}
  \par
   \vspace{2mm} 
\noindent
In this appendix we investigate the various integrability conditions
which appeared in section 3 of this paper, see also \cite{yan}.
{}From two given $(1,1)$ tensors $R$ and $S$, we can construct a $(1,2)$
tensor $N[A,B]$, the Nijenhuis tensor:
\begin{eqnarray}
N[R,S](U,V)=\frac 1 2 \left(
[RU,SV]-R[U,SV]-S[RU,V]+RS[U,V]+ R\leftrightarrow S\right).
\end{eqnarray}
For 3 $(1,1)$ tensors $R$, $S$ and $T$, one can verify by direct computation
the following identity discovered by Frochlicher and Nijenhuis:
\begin{eqnarray}
&&{\cal T }[R,S,T](U,V)\equiv
N[R,ST](U,V)+N[S,RT](U,V)-R\,N[S,T](U,V)-\nonumber\\
&&\qquad\qquad S\,N[R,T](U,V)
-N[R,S](TU,V)-N[R,S](U,TV)=0.
\end{eqnarray}
If we have two {\it commuting} $(1,1)$ tensors $R$ and $S$, we can
construct another $(1,2)$ tensor:
\begin{eqnarray}
M[R,S](U,V)=
[RU,SV]-R[U,SV]-S[RU,V]+RS[U,V].
\end{eqnarray}
In particular we get that
\begin{eqnarray}
N[R,S](U,V)=\frac 1 2 \left( M[R,S](U,V) + M[S,R](U,V)\right).
\end{eqnarray}
Given three mutually commuting $(1,1)$ tensors  $R$, $S$ and $T$, one obtains
through direct computation the identity
\begin{eqnarray}
{\cal U }[R,S,T](U,V)\equiv
M[R,ST](U,V)- S\,M[R,T](U,V)
-M[R,S](U,TV)=0.
\end{eqnarray}
We now prove the following lemma:

\vspace{0.7cm}

\noindent{\it \underline{Lemma 1}:
Given $J$ and $\bar J$, two complex structures which commute, $[J,
\bar J]=0$, then $N[J,J](U,V)=N[\bar J, \bar J](U,V)=0$ imply that
$N[\Pi,\Pi](U,V)=N[\Pi,J](U,V)=N[\Pi,\bar J](U,V)=N[J,\bar J](U,V)=0$ with
$\Pi\equiv J\bar J$.}

\vspace{0.7cm}

\noindent Using the definition of the Nijenhuis tensor, one gets
\begin{eqnarray}
&&N[J,J](\Pi U,V)+N[J,J](U,\Pi V)+\Pi N[J,J](U,V)-\nonumber\\
&&\qquad\qquad\Pi N[\bar J,\bar J](\Pi
U, \Pi V)=\Pi N[\Pi,\Pi](U,V)-N[J,\bar J](U,V).
\end{eqnarray}
Using  $N[J,J](U,V)=N[\bar J,
\bar J](U,V)=0$, we get from this:
\begin{eqnarray}
\Pi N[\Pi,\Pi](U,V)=N[J,\bar J](U,V). \label{a1}
\end{eqnarray}
{}From now on we continously use $[J,\bar J]=0$ and $N[J,J](U,V)=N[\bar J,
\bar J](U,V)=0$. From ${\cal T}[J,J,\Pi](U,V)=0$, we get
\begin{eqnarray}
N[J,\Pi](U,V)=J\,N[J,\bar J](U,V), \label{a2}
\end{eqnarray}
and from ${\cal T}[\bar J,\bar J,\Pi](U,V)=0$,
\begin{eqnarray}
N[\bar J,\Pi](U,V)=\bar J\,N[ J,\bar J](U,V).  \label{a3}
\end{eqnarray}
Using this in ${\cal T}[J,\bar J,\bar J](U,V)=0$ yields,
\begin{eqnarray}
N[J,\bar J](\bar J U, V)+ N[J,\bar J](U,\bar J V)=0. \label{a4}
\end{eqnarray}
Using eqs. (\ref{a1}), (\ref{a2}) and (\ref{a4}) in ${\cal T}[J,\Pi,\bar
J](U,V)=0$, gives
\begin{eqnarray}
N[J,\bar J](U,V)=0.
\end{eqnarray}
Combining this with eqs. (\ref{a1}-\ref{a3}) shows that also all other
Nijenhuistensors vanish. Which proves the lemma.
As the complex structures commute, this results in
additional $(1,2)$ tensors which can be constructed out of $J$ and $\bar J$:
$M[J,\bar J](U,V)$,
$M[\Pi, J](U,V)$ and  $M[\Pi,\bar J](U,V)$. We now prove an additional lemma:

\vspace{0.7cm}

\noindent{\it \underline{Lemma 2}: For $J$ and $\bar J$ two commuting complex
structures,
we have that
$N[J,J](U,V)$ $ = N[\bar J, \bar J](U,V)=0$ imply the vanishing of the
tensors
$M[J,\bar J](U,V)$,
$M[\Pi, J](U,V)$ and  $M[\Pi,\bar J](U,V)$.}

\vspace{0.7cm}

\noindent We use throughout the previous lemma.
{}From ${\cal U}(\bar J,\bar J,J)=0$ we get
\begin{eqnarray}
M[\bar J,\Pi](U,V)=\bar J M[\bar J,J](U,V),\label{b1}
\end{eqnarray}
${\cal U}(\bar J,J,\bar J)=0$ implies
\begin{eqnarray}
M[\bar J,\Pi](U,V)= M[\bar J,J](U,\bar J V),\label{b2}
\end{eqnarray}
and from  ${\cal U}(J,\bar J,\bar J)=0$ we obtain:
\begin{eqnarray}
M[\bar J,J](U,\bar J V) =-\bar J M[\bar J,J](U, V).\label{b3}
\end{eqnarray}
Using eq. (\ref{b3}) in eq. (\ref{b2}) yields
\begin{eqnarray}
M[\bar J,\Pi](U,V)=-\bar J M[\bar J,J](U,V)
\end{eqnarray}
Comparing this with eq. (\ref{b1}) gives $M[\bar J,\Pi](U,V) =
M[\bar J,J](U,V)=0$. Finally, using this result in ${\cal U}(J,J,\bar
J)=0$ results in the vanishing of $M[ J,\Pi](U,V)$, which proves the lemma.

\end{document}